\documentclass[twocolumn,aps,showpacs,superscriptaddress,pre]{revtex4}

\newcommand{\re}{\rho_f}  
\newcommand{\rt}{\rho_t}  
\newcommand{\fa}{{\mathfrak n}}  

\usepackage{amsfonts}
\usepackage{amsmath,amssymb,graphicx}
\usepackage{color}

\begin{document}
\title{Population Annealing Simulations of a Binary Hard Sphere Mixture}

\author{Jared Callaham}
\email{jared.callaham@gmail.com}
\affiliation{Department of Physics, University of Massachusetts,
Amherst, Massachusetts 01003 USA}

\author{Jonathan Machta}
\email{machta@physics.umass.edu}
\affiliation{Department of Physics, University of Massachusetts,
Amherst, Massachusetts 01003 USA}
\affiliation{Santa Fe Institute, 1399 Hyde Park Road, Santa Fe, New 
Mexico 87501 USA}

\begin{abstract}

Population annealing is a sequential Monte Carlo scheme well-suited to simulating equilibrium states of systems with rough free energy landscapes.  Here we use population annealing to study a binary mixture of hard spheres.  Population annealing is a parallel version of simulated annealing with an extra resampling step that ensures that a population of replicas of the system represents the equilibrium ensemble at every packing fraction in an annealing schedule.  The algorithm and its equilibration properties are described and results are presented for a glass-forming fluid composed of a 50/50 mixture of hard spheres with diameter ratio of 1.4:1.  For this system, we obtain precise results for the  equation of state in the glassy regime up to packing fractions $\varphi \approx 0.60$ and study deviations from the BMCSL equation of state.  For higher packing fractions, the algorithm falls out of equilibrium and a free volume fit predicts jamming at packing fraction $\varphi \approx 0.667$.  We conclude that population annealing is an effective tool for studying equilibrium glassy fluids and the jamming transition.

\end{abstract}
\maketitle

\section{Introduction}
\label{sec:intro}

One of the grand challenges of computational statistical physics is to understand the nature of glassy systems.  Profound open questions remain concerning  both configurational glasses and spin glasses.  The signature property of glassy systems is an extreme slowing of dynamics in both laboratory and computational experiments.  The equilibrium properties of glassy systems are thus very difficult to study.  The situation is somewhat better for spin glasses where it is established that there is thermodynamic glass transition and a well-understand mean field theory of the low temperature glass phase, though controversies surround the nature of the low temperature phase in finite dimensions.   For configurational glasses it is not even known whether there is an underlying thermodynamic transition or whether the glass transition is entirely a kinetic phenomenon.   

Much of the progress for finite-dimensional spin glasses has been made possible due to algorithmic advances, particularly the introduction of replica exchange Monte Carlo, also known as parallel tempering \cite{SwWa86,HuNe96,KaPaYo01,BaCrFe10,YuKaMa12}.   The situation for simulating equilibrium glassy fluids is less well-developed and fundamentally a more difficult problem.  Parallel tempering has also been extensively applied to fluid systems~\cite{Okabe2001435} in the glassy regime~\cite{Odriozola2009,Odriozola2011}.  Other recent algorithmic advances such as event chain Monte Carlo \cite{Bernard2009,Isobe2015} and particle-swap Monte Carlo \cite{Berthier2016} have also shown promises.  

In this paper we introduce population annealing Monte Carlo as a method for studying fluid systems and show that it is an effective tool for studying  equilibrium properties up to high densities in the glassy regime.  Population annealing was first developed for spin glasses~\cite{HuIb03,Machta2010,Wang2015} and has been shown to be an efficient method for large-scale studies of equilibrium states~\cite{WaMaKa14,WaMaKa15a} and ground states~\cite{WaMaKa15} of spin glasses.
Here we use population annealing Monte Carlo to simulate a binary mixture of hard spheres.  

In particular, we study a glass-forming 50/50 binary mixture of hard spheres with a diameter ratio of 1.4:1, which has been the subject of  previous computational studies, e.g.\ \cite{Berthier2009, Brambilla09, Odriozola2011}.    Using population annealing we obtain precise estimates of the equation of state up to packing fractions in the glassy regime.  We compare our results to previous simulations and the well-known Boublik-Mansoori-Carnahan-Starling-Leland equation of state~\cite{Boublik1970,Mansoori1971}.

The paper is organized as follows.  In Sec.\ \ref{sec:pa} we introduce population annealing and describe its properties. In Sec.\ \ref{sec:mome} we describe the hard sphere fluid model, the simulation methods and the observables that we study.  Results for the equation of state, the approach to jamming and the performance of the algorithm algorithm are presented in Sec.\ \ref{sec:results}.  The paper concludes with a discussion.

\section{Population Annealing Monte Carlo}
\label{sec:pa}

\subsection{Overview}

Population annealing (PA) is closely related to simulated annealing.  In both algorithms, a system is taken through an annealing schedule in one or more thermodynamic control parameters, e.g. temperature or density, from a region where equilibration is easy, e.g. high temperature or low density, to a region where equilibration is difficult.  At each step in the annealing schedule, an equilibrating procedure, e.g. the Metropolis algorithm, is applied to the system at the current value of the control parameter(s).  The objective of simulated annealing is to find ground states or low-lying states and simulated annealing does not  sample from the equilibrium distribution at each step along the annealing schedule.  Population annealing, by contrast, does sample from the equilibrium distribution at each step along the annealing schedule.  In PA a large population of replicas of the system is annealed and, at each step in the annealing schedule,  the population is resampled to maintain equilibrium as described in the next section.

\subsection{Population Annealing for Hard Spheres}

In this section we describe population annealing for hard sphere fluid systems using packing fraction $\varphi$ as the control parameter.  A population of $R$ independent replicas of the system is initially prepared at some low (or zero) packing fraction, $\varphi_0$.   In the simulations reported below $R = 10^6$ and $\varphi_0=0$.  In parallel, each replica is taken through an annealing schedule, which is a sequence of increasing packing fractions $\{ \varphi_0, \varphi_1, \ldots, \varphi_K\}$.  Each annealing step $(\varphi_i \rightarrow \varphi_{i+1})$ consists of two parts.  In the first part, an equilibrating procedure is applied at the initial density, $\varphi_i$.  In our study, the equilibrating procedure is event chain Monte Carlo (ECMC) \cite{Bernard2009} (see Sec.\ \ref{sec:eventchain}) but other methods such as molecular dynamics or Metropolis Monte Carlo would also be suitable.  The second part of the annealing step is to increase the packing fraction from $\varphi_i$ to $\varphi_{i+1}$ holding the relative positions of the particles fixed.   Typically some replicas in the population will now suffer overlaps between spheres and  therefore have disallowed configurations.  These disallowed replicas are removed from the population.  Suppose that $\epsilon R$ replicas are culled from the population.  The population size is restored to $R$ by randomly choosing $\epsilon R$ replicas from among the surviving $(1-\epsilon)R$ replicas and making one copy of each of them.   For an observable ${\cal O}$, the PA estimator, $\tilde{\cal O}$ for the ensemble average of an observable, $\langle {\cal O} \rangle$ at each packing fraction is obtained from an average over the population at that packing fraction.  The population can be thought of as an approximate, finite realization of the statistical ensemble for hard spheres.  Note that it is possible that there are ${\it no}$ allowed configurations at the new packing fraction.  In this case, the algorithm must be terminated without producing results for higher packing fractions.

In the limit of large $R$, if the original population  is an unbiased sample from the equilibrium hard sphere ensemble at packing fraction $\varphi_i$ then the new population will also be an unbiased sample from the equilibrium hard sphere ensemble at packing fraction $\varphi_{i+1}$.  However, the new population is now correlated due to the copying of replicas and, for finite $R$, the new population may be biased due to the omission of configurations that are important at the higher density but too rare to be represented in the population at the lower density.  The equilibrating procedure at the new density partially corrects these problems.  An analysis of the error in PA due to finite population size is given in Sec.\ \ref{sec:error}.

Population annealing may be implemented with a fixed annealing schedule however we have found it more convenient to use an adaptive annealing schedule in which the fraction culled in each step is fixed.  In our simulations the culling fraction is $\epsilon = 0.05$.  In this implementation, the annealing schedule is a random list is a random list of packing fractions.  However, in practice spacing between successive values of $\varphi$ is very small so it is straightforward to interpolate to obtain observables at any packing fraction.  For the adaptive annealing schedule, in principle, the algorithm will never terminate but it may jam in the sense of taking smaller and smaller steps converging to a maximal or jammed density. 

Population annealing for hard spheres in the NVT ensemble is particularly simple because each allowed hard sphere configuration has the same probability in the equilibrium ensemble so the resampling step requires no re-weighting or Boltzmann factors.  It would also be possible to use  the NPT ensemble (with fixed temperature), in which case a Boltzmann factor $\exp[-\beta(p_{i+1} - p_{i}) V_r]$ would be required when deciding how many copies to make of replica $r$ in the step  $(p_i \rightarrow p_{i+1})$ where the volume of replica $r$ at pressure $p_i$ is $V_r$.  

\subsection{Configurational entropy estimator}
One desirable feature of PA is that it gives direct access to thermodynamic potentials.  For the hard sphere version of the algorithm described here, we have direct access to the configurational entropy $S_c(\varphi)$ as a function of packing fraction.  From the basic definition of entropy, we have $S_c =  \log \Omega$ where $\Omega$ is the statistical weight or dimensionless volume in configuration space of accessible configurations and the units are chosen so that Boltzmann's constant is unity.  The factor $1-\epsilon$ is an estimator of the ratio of the statistical weight before and after the culling, $\Omega_{i+1}/\Omega_i$,  during the annealing step $(\varphi_i \rightarrow \varphi_{i+1})$.  Thus we have the following expression for the change in the estimator $\tilde{S_c}$ in one annealing step,
\begin{equation}
\label{eq:iterateS}
\tilde{S_c}(\varphi_{i+1}) - \tilde{S_c}(\varphi_{i}) \approx  \log (1- \epsilon) .
\end{equation} 
Summing this estimate over the annealing schedule up to step $k$, we obtain the entropy at packing fraction $\varphi_k$ in terms of the entropy at the initial packing fraction and the number of annealing steps taken,
\begin{equation}
\label{eq:entropy}
\tilde{S_c}(\varphi_k) = \tilde{S_c}(\varphi_0) +  k \log (1-\epsilon) .
\end{equation}
The equation of state can be obtained by differentiating with respect to $\varphi$.  These estimates become exact in the $R \rightarrow \infty$ limit.  Errors are discussed in Sec.\ \ref{sec:error}.

\subsection{Weighted Averaging}
A very useful feature of PA is the ability to improve both statistical and systematic errors by combining many independent runs using weighted averaging.  If the population is perfectly equilibrated and there are no systematic errors then the most efficient way to reduce statistical errors in an observable is to perform ordinary, unweighted averaging over the values of the observable in each run.  However, for finite $R$, each run is not completely equilibrated and there are systematic errors.  Unweighted averaging suppresses statistical errors but does not suppress systematic errors.  However, using weighted averaging we can also suppress systematic errors.   Suppose we have measured both the configurational entropy estimator, $\tilde{S_c}$ and  the estimator of an observable $\tilde{\cal O}$ at a given value of $\varphi$ in each of $M$ independent runs, all with the same population size.     The weighted average for the observable, $\overline{{\cal O}}$ is  
\begin{equation}
\label{eq:weight}
\overline{{\cal O}} = \frac{\sum_{m=1}^M   \tilde{\cal O}^{(m)}  \exp[\tilde{S_c}^{(m)}] } {\sum_{m=1}^M   \exp [\tilde{S_c}^{(m)} ]},
\end{equation}
where $m$ indexes the independent runs.  The observable ${\cal O}$ must be a quantity that can be measured in a single replica at a single packing fraction.  Examples include pressure and various correlation functions but not overlaps between replicas or the entropy itself.

We make the following claim:
Weighted averaging, as defined in Eq.\ \eqref{eq:weight},  yields an exact, unbiased result for fixed population size $R$, and fixed annealing schedule $\{ \varphi_0, \varphi_1, \ldots, \varphi_K\}$, in the limit of infinitely many runs, $M \rightarrow \infty$.  The same conclusion holds for the adaptive annealing schedule used here and is discussed below. 

The validity of weighted averaging is trivial for $R=1$, (simulated annealing).  For a given $\varphi_i$, a run either survives to that packing fraction or is terminated at a lower packing fraction.  Surviving runs all have the same weight since there is no resampling, while terminated runs have zero weight.  Each surviving run is an unbiased sample taken from the hard sphere ensemble.  Since the weighting factor is either 0 or 1, the weighted average is a simple average over the surviving runs.  The $R=1$ version of the algorithm lacks resampling and is therefore highly inefficient at high densities.

The validity of weighted averaging for arbitrary $R$ can be established using an inductive argument.  The populations of the runs at the initial packing fraction $\varphi_0$ are assumed to be equilibrated so that unweighted averaging is appropriate at $\varphi_0$.  Correspondingly, the weight factors are the same  and the claim is trivially valid for $\varphi_0$.  Now suppose that the weighted averaging claim holds at packing fraction $\varphi_i$ with weights $w_i^{(m)}$.  In the first part of the  resampling step from $\varphi_{i}$ to $\varphi_{i+1}$, the random fraction $\epsilon_i^{(m)}$ of replicas with hard sphere overlaps are culled from the population.  Consider the situation before copying is done to restore the population size to $R$.  Each surviving replica in run $m$ should still have the weight $w_i^{(m)}$ if one averages over all replicas in all runs.  However, what we actually do is to first average the observable within each run and then average over runs.  Therefore weight of each run must be adjusted to reflect its new population size,
\begin{equation}
\label{ }
w_{i+1}^{(m)} \propto  (1-\epsilon_i^{(m)})w_i^{(m)},
\end{equation} 
where the constant of proportionality is set by the requirement that the weights are normalized.  Comparing to Eq.\ \eqref{eq:iterateS}, we see that if $w_i^{(m)} \propto \exp[ \tilde{S_c}^{(m)}(\varphi_i)]$ then $w_{i+1}^{(m)} \propto \exp[ \tilde{S_c}^{(m)}(\varphi_{i+1})]$ and the inductive hypothesis is verified.  The remainder of the annealing step $(\varphi_i \rightarrow \varphi_{i+1})$ consists of randomly copying replicas to restore the population size to $R$ and then applying the equilibrating procedure to each replica.  The copying step yields additional fluctuations but does not change the expectation of the observable for each run.  The equilibrating procedure may change the expectation of the observable for a given run if the individual runs were initially out of equilibrium.  Nonetheless, the correctness of weighted averaging  is preserved because the equilibrating procedure  and  weighted averaging both converge to the same hard sphere distribution.   

Weighted averaging for the entropy itself differs from other observables because the entropy is obtained from a sum over all packing fractions.  Nonetheless, the final result for the weighted average of the entropy $\overline{S_c}$ takes a simple form,
\begin{equation}
\label{eq:weightedS}
\overline{S_c} = \log \frac{1}{M} \sum_{m=1}^M \exp[\tilde{S_c}^{(m)}] .
\end{equation}
This result is an example of the Jarzynski equality~\cite{Jarz97}.  A derivation for PA in the the context of free energy rather than the entropy can be found in Ref.\ \cite{Machta2010}.

Finally, in the adaptive step algorithm used here, the annealing schedule is a random list so each run visits a distinct set of packing fractions whereas it is necessary to carry out weighted averages at fixed values of $\varphi$.   To accomplish this we interpolate both the observables and the entropy between packing fractions and then use the weighted averaging formulas, Eqns.\ \eqref{eq:weight} or \eqref{eq:weightedS}.

\subsection{Equilibration and Systematic Errors}
\label{sec:error}
How do we know whether PA is yielding equilibrium values of observables?  As discussed above, the resampling step for  finite population size introduces systematic errors because the distribution at the new packing fraction is not fully sampled.  The exactness of weighted averaging gives a way to quantify the deviations from equilibrium.  The analysis of systematic errors is discussed in detail in Ref.\ \cite{Machta2010} and is briefly reviewed here.

The expected value of an observable from a single run at population size $R$ of PA is the {\em unweighted} average over runs while the the exact value of the observable is the {\em weighted} average (both averages taken in the limit of infinitely many runs).  Thus the systematic error in an observable,  $\Delta {\cal O}$  is given by $\Delta {\cal O} =  \langle \tilde{\cal O} \rangle-  \overline{\cal O}$ where $\langle \tilde{\cal O} \rangle$ represents an unweighted average over independent runs of the algorithm.  The difference between weighted and unweighted averages depends on the variance of the weighting factor in Eq.\ \eqref{eq:weight}.   For the case that the joint distribution of $\tilde{\cal O}$ and $\tilde{S_c}$ is a bivariate Gaussian, it  was shown in Ref.\ \cite{Wang2015} that systematic errors  in  measuring ${\cal O}$ are given by the covariance of $\tilde{\cal O}$ and $\tilde{S_c}$,
\begin{equation}
\label{eq:syserror}
\Delta {\cal O} =  {\rm cov}(\tilde{\cal O},\tilde{S_c}) = {\rm var}(\tilde{S_c}) \left[\frac{{\rm cov}(\tilde{\cal O},\tilde{S_c})}{{\rm var}(\tilde{S_c})}\right].
\end{equation}  
The second, trivial identity in Eq.\ \ref{eq:syserror} is useful because the ratio in the square brackets goes to a constant that depends on ${\cal O}$ as $R \rightarrow \infty$ while ${\rm var}(\tilde{S_c})$ diminishes as $1/R$ and is independent of the observable.  
The expression on the far right in Eq.\ \eqref{eq:syserror} motivates defining an equilibration population size $\rho_f$,
\begin{equation}
\label{eq:re}
\re = \lim_{R \rightarrow \infty} R \ {\rm var}(\tilde{S_c}).
\end{equation}
Systematic errors in all observables are proportional to $\re/R$ and PA simulations are well-equilibrated when $R \gg \re$.   In population annealing $\re$ plays the same role as the exponential autocorrelation time in Markov chain Monte Carlo methods.  

For sufficiently large $R$,  $\tilde{S}_c$ and other observables arise from many independent, additive contributions so non-rigorous ``central limit theorem" arguments suggest that the joint distribution of entropy and any observable should be a bi-variate Gaussian.  The approach to Gaussianity with $R$ was investigated in Ref.\ \cite{Wang2015}.  If $R$ is too small the joint distribution will not be Gaussian and there will be corrections to Eq.\ \eqref{eq:syserror} containing higher cumulants of the joint distribution.   We shall see later that for the fixed population size used here, the joint distribution of pressure and entropy is a bivariate Gaussian for small packing fractions but significant exponential tails appear at high packing fractions.  These corrections and their meaning are explored in Sec.\ \ref{sec:paperf}

In our simulations we use weighted averaging extensively.  What are the systematic errors in weighted averages?  The same arguments that shows that  ${\rm var}(\tilde{S_c})$ controls systematic errors in a single run demonstrate that the relevant measure of equilibration for a weighted average is ${\rm var}(\overline{S_c})$, where $\overline{S_c}$ is the weighted average estimator of the configurational entropy defined in Eq.\ \eqref{eq:weightedS}.    From ${\rm var}(\overline{S_c})$ we define the weighted average equilibration population  size $\re^*(R)$ that is a function of the fixed population size $R$ used in the weighted average,
\begin{equation}
\label{eq:re-star}
\re^*(R) = \lim_{M \rightarrow \infty} M R \; {\rm var}(\overline{S_c}).
\end{equation}
Since weighted averaging using $M$ runs of population size $R$ is less efficient than doing a single large run with population size $MR$, we have that  $\re^*(R) \geq \re$ and we also expect that $\lim_{R \rightarrow \infty} \re^*(R) \rightarrow \re$.

Of course, it is not practical to compute ${\rm var}(\overline{S_c})$ from repeated experiments each with $M$ runs.  Instead we use a bootstraps procedure to estimate ${\rm var}(\overline{S_c})$ by resampling with replacement from our $M$ runs.  Each sample of size $M$ is employed to compute $\overline{S_c}$ from Eq.\ \eqref{eq:weightedS} and the variance is compute by resampling many times.

It is known that in the thermodynamic limit and at sufficiently high packing fraction the equilibrium state of the binary mixture studied here is phase separated with separate crystals of the small and large spheres~\cite{Hopkins2012}.  For a finite number of spheres the dominant configurations in the equilibrium distribution are not known since the entropic cost of the interface between the two crystalline regions may prevent phase separation.  It may be that the equilibrium states are dominated either by phase separated configurations or other non-random configurations.  We have not seen evidence of either phase separation or other forms of ordering in our PA simulations.  

%

\subsection{Statistical Errors}
\label{sec:errort}
If an observable ${\cal O}$ is averaged over $R$ independent measurements, the statistical error $\delta {\cal O}$ in the mean is given by $\sqrt{{\rm var}({\cal O})/R}$.  The resampling step in PA introduces correlations so that statistical errors are larger than $\sqrt{{\rm var}({\cal O})/R}$.  As discussed in detail in Ref.\ \cite{Wang2015}, we can bound statistical errors by ignoring the de-correlating effects of the equilibrating procedure.  If there is no equilibrating procedure, each descendent of an initial replica is the same.  We refer to the set of descendants of an initial replica as a {\it family} and define $\fa_i$ to be the fraction of the population in family $i$.  In the absence of the equilibrating procedure ${\cal O}$ takes a single value for every member of the family, call this value ${\cal O}_i$.  Given this assumption,
\begin{equation}
\label{ }
\tilde{\cal O} = \sum_i {\cal O}_i \fa_i.
\end{equation}
To proceed we make two additional assumptions: (1)  the observable ${\cal O}_i$ and family fraction $\fa_i$ are uncorrelated and (2) the distribution of family fraction $\fa_i$ has small fluctuations from run to run for large $R$. Given these assumptions, the variance of the observable is  bounded by,
\begin{equation}
{\rm var}(\tilde{\cal O}) \leq {\rm var}({\cal O}) \sum_i  \fa_i^2 .
\end{equation}

Assuming that the second moment of the family fraction 
scales as $1/R$ we define the {\em mean square family
size}, $\rt$,
\begin{equation}
\label{eq:rt}
\rt = \lim_{R \rightarrow \infty} R \ \sum_i  \fa_i^2.
\end{equation}
The bound on the statistical error in $\delta
\tilde{\cal O}$ becomes
\begin{equation}
\label{eq:staterror}
\delta \tilde{\cal O} \leq \sqrt{{\rm var}({\cal O})\rt /R}.
\end{equation}
Note that if there is no decimation so that $\fa_i = 1/R$, then $\rt=1$ and Eq.\ \eqref{eq:staterror} (as an equality) reduces to the expression for statistical errors for uncorrelated measurements.  Since $\rt$ provides only a bound on statistical errors, in practice we use bootstrapping to estimate errors. 

In Ref.\ \cite{toappear} it is shown that $\re$ and $\rt$ are close to one another.  We investigate the relation between these two measures in Sec.\ \ref{sec:paperf}.  Since measuring $\re$ requires multiple runs while $\rt$ can be measured from a single run, $\rt$ can serve as a practical measure of equilibration.

 It is important to note that in the regime that $R \gg \rho_f$, systematic errors diminish as $1/R$ while statistical errors diminish as $1/\sqrt{R}$ so that, generally, systematic errors are much smaller than statistical errors.
 
\subsection{Comparison to Parallel Tempering}

Population annealing is a sequential Monte Carlo method \cite{DoFrGo01} in contrast to most simulation methods in statistical physics, which are Markov chain Monte Carlo methods.  In sequential Monte Carlo, equilibration is approached as the population size $R$ increases while for Markov Chain Monte Carlo, equilibration is approached as the number of Monte Carlo sweeps is increased.  Among Markov chain Monte Carlo methods parallel tempering shares the greatest similarity with population annealing.  It is comparably efficient to parallel tempering but, as we have seen, it has some interesting advantages--it gives direct access to thermodynamic potentials such as entropy or free energy, it is amenable to massive parallelization and multiple runs can be combined to improve both statistical {\em and} systematic errors.   Parallel tempering explores the disconnected minima in a rough free energy landscapes by annealing to low temperatures multiple times, using correctly designed swap moves to insure that each region of the free energy landscape is visited with the statistical weight determined by the Gibbs distribution.  Population annealing explores rough free energy landscapes in a single annealing run using a large number of replicas to populate the disconnected minima in the landscape.  Resampling insures that each minimum contains a fraction of the population given by the Gibbs distribution.

For parallel tempering and other Markov chain Monte Carlo methods, systematic and statistical errors can be determined from autocorrelation functions of observables.  Systematic errors are related to the exponential autocorrelation time while statistical errors are related to the integrated autocorrelation time.  Thus the two characteristic population sizes, $\re$ and $\rt$, play analogous roles for PA to the exponential and integrated autocorrelation times, respectively.

\section{Model and Methods}
\label{sec:mome}

\subsection{Hard Sphere Model}
\label{sec:model}
Hard spheres have long offered a simple and useful model for studying the properties of fluids. Although these systems have been studied for decades, there are still mysteries surrounding some of their behaviors. Monodisperse systems exhibit a distinct entropy-driven first-order phase transition from the disordered liquid to a crystalline solid, but a metastable fluid branch persists beyond the transition and its high-density behavior is still not fully understood. It is as yet unclear to what packing fraction this metastable branch persists~\cite{Berthier2009, Brambilla09, Odriozola2009, Berthier2016}, and there has also been evidence for~\cite{Berthier2009, Hermes2010} and against~\cite{Odriozola2011, Berthier2016} the existence of a thermodynamic glass transition.

One difficulty in studying the high-density behavior of the metastable fluid is that, since the true thermodynamic state is a solid, full equilibration will lead to crystallization. Size polydispersity is often introduced to avoid crystallization.  We study 50/50 mixture of spheres with diameter ratio 1.4:1, which does not easily crystallize~\cite{Ohern2002, Berthier2009, Hopkins2012}.

At low and moderate densities the equation of state of a binary mixture of hard spheres is accurately described by the Boubl\'ik-Mansoori-Carnahan-Starling-Leland (BMCSL) equation\cite{Boublik1970, Mansoori1971}: 
\begin{equation} \label{eq:bmcsl}
Z(\varphi) = \frac{ (1 + \varphi + \varphi^2) - 3\varphi(y_1 + y_2\varphi) - y_3\varphi^3}{(1-\varphi)^3}
\end{equation}
The constants $ y_1 $, $ y_2 $, and $ y_3 $ depend on the choice of polydispersity and for the 50/50 binary mixture with polydispersity ratio 1.4:1 they are given by, $y_1 = 0.0513$, $y_2 = 0.0237$ and $y_3 = 0.9251$.

\subsection{Event Chain Monte Carlo}
\label{sec:eventchain}

Population annealing requires an equilibrating procedure performed at each annealing step. We use  event chain Monte Carlo (ECMC), which has been shown to be highly efficient for simulating 2D and 3D hard sphere systems \cite{Engel2013, Isobe2015, Michel2014, Bernard2009, Bernard2011, Bernard2011a}.

In an ECMC step, a particle is chosen at random and moved in a given direction. The particle moves until it collides
with some other particle (the ``€œevent''€). The original particle remains at the point of collision, while the particle that was struck
then moves in the same direction. This process continues until the total displacement length of the chain of particles reaches a predetermined chain length $ \ell_c $. The chain length is a parameter that can be adjusted, e.g. to minimize correlation times. 

There are several  variants of ECMC, but the simplest and 
fastest \cite{Bernard2009} version has only two directions of motion, $+x$ or $+y$. This method, called ``x-y straight event chain,''  violates detailed balance, but preserves global balance and thus converges to the correct equilibrium state \cite{Bernard2009}.

\subsection{Population Annealing Pseudo-Code}

The following is pseudocode for our implementation of population annealing with event chain Monte Carlo for a polydisperse hard sphere system. The diameters of the small and large particles are fixed respectively at 1 and 1.4.

\begin{figure}[htbp] 
   \centering
   \includegraphics[width=0.5\linewidth]{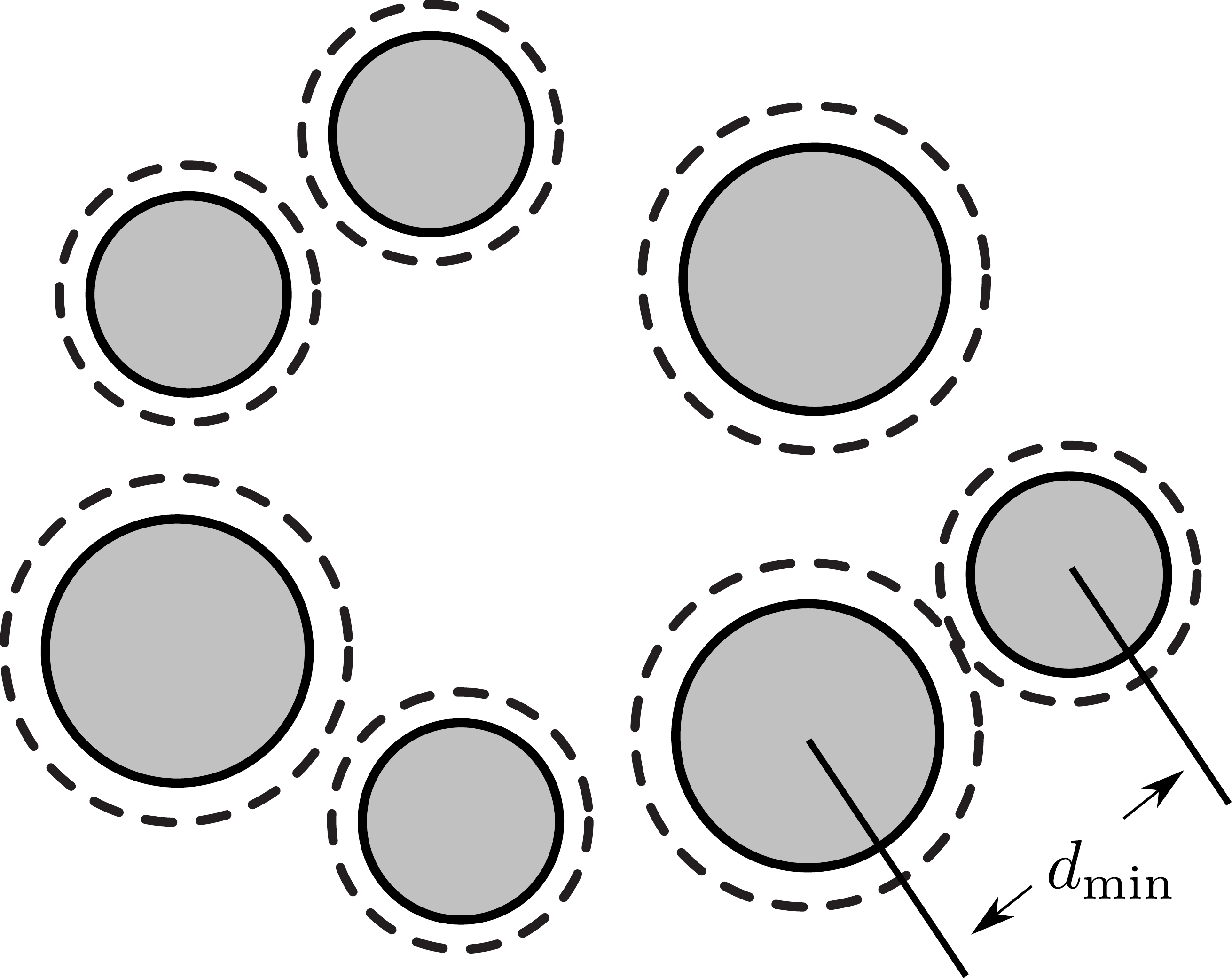} 
   \caption{The minimum relative center-to-center distance over all particle pairs in a replica is $ d_\mathrm{min}$.}
   \label{fig:compression-step}
\end{figure}

\begin{enumerate}
\item Initialize each of $R$ replicas at zero packing fraction by choosing $N$ points at random within a unit cube simulation cell with periodic boundary conditions. Each particle is also labeled with a size according the prescription for polydispersity.

\item For each replica, determine $ d_\text{min} $, the minimum relative center-to-center distance between any two particles as shown in Fig. \ref{fig:compression-step}: \begin{equation}
d_\text{min} = \min_{i, j} \bigg[ \frac{2 \, |\bf{r_i} - \bf{r_j}|}{\sigma_i + \sigma_j} \bigg],
\end{equation} where $ \bf{r_i} $ is the position of particle $ i $ and $ \sigma_i $ is its diameter. The dimensions of this system and particle positions may be rescaled by $ 1/d_\text{min} $ without causing overlaps. Note that for the first annealing step this rescaling increases the size of the system, and for all later steps the size is decreased.

\item Sort the list of $ d_\text{min} $ values and define the $ (\epsilon R)^{\text{th}} $ smallest value of $d_\text{min} $ as $ d^*$.  

\item Rescale the simulation box size $L$, packing fraction $\varphi$ and position $\bf{r} $ of each particle according to $ L \leftarrow L/d^* $, $ \varphi \leftarrow \varphi/(d^*)^3 $ and $ {\bf r} \leftarrow {\bf r}/d^* $.  (Note that $ \varphi $ is initially zero and needs to be calculated after the first annealing step).

\item After rescaling there are $ \epsilon R $ replicas with invalid configurations \footnote{Note that this choice of rescaling introduces a tiny deviation from the hard sphere distribution because one pair of particles out in one replica is {\em always} in contact immediately after the rescaling.  The resulting error is far smaller than other systematic and statistical errors in the simulation.}.  These replicas are eliminated from the population. The same number of  replicas are randomly chosen (with replacement) from among the valid replicas and copies of these replicas are added to the population. After this step, the population consists of $ R $ valid configurations.

\item  Perform one ECMC sweep on each replica. A sweep should displace on average $ N $ particles. We choose the chain length $ \ell_c $ to displace an average of $ \sqrt{N} $ particles (See Appendix \ref{app:chain-length}) and then perform $ \sqrt{N} $ ECMC moves.

\item Repeat steps 2-6 on each replica until a predefined packing fraction is reached.

\end{enumerate}

\subsection{Observables}
\label{sec:observables}

\subsubsection{Entropy}
The relative configurational entropy $\tilde{S}_c$ at each step in the annealing schedule is obtained from Eq.\ \eqref{eq:entropy}.  The simulations are initialized at $\varphi_0=0$ and the entropy at zero packing fraction is taken to be zero so that all entropy values are negative.  We would like to know the weighted average value of the entropy $\bar{S}_c$ as a function of $\varphi$.  However, our annealing schedule traverses a fixed set of entropies and the packing fraction at each entropy is variable.  Values of the entropy at each packing fraction are obtained by interpolation and Eq.\ \eqref{eq:weightedS} is applied to obtain the weighted average entropy.

\subsubsection{Equation of State}
The dimensionless equation of state, $Z$ is expressed in terms of the pressure $P$, 
\begin{equation}
Z = \frac{\pi \sigma^3}{6} \frac{P}{\varphi},
\end{equation} 
where $\sigma^3$ is the average cubed sphere diameter, which for a binary mixture with diameters $\sigma_A$ and $\sigma_B$ is given by  $\sigma^3 = \frac{1}{2} (\sigma_A^3 + \sigma_B^3)$.    We obtain $Z$ at each annealing step and then use interpolation to determine $Z$ at a fixed set of packing fractions.  We report the weighted average equation of state $\bar{Z}$ as a function of packing fraction. Error bars are obtained by bootstrapping the weighted averages.   Two independent ways of measuring the pressure are described in the next two subsections.

\subsubsection{Thermodynamic Pressure Estimator}
Using Eq.\ \eqref{eq:entropy}, pressure and equation of state can be found through the thermodynamic definition,
 \begin{equation} 
 \frac{P}{T} = \frac{\partial S}{\partial V} .
 \end{equation}
In our simulations the change in entropy for each annealing step is a constant $\log(1-\epsilon)$ and the increment in packing fraction is variable. Thus a thermodynamic estimator for the pressure at annealing step $ k $ is 
\begin{equation} \label{eq:pa-pressure}
\frac{P_k}{T} = -\frac{\varphi_k}{V_k} \frac{\Delta S}{\Delta \varphi_k} = -\frac{\varphi_k}{V_k} \frac{\log (1-\epsilon)}{\Delta \varphi_k}.
\end{equation} Setting $ T = 1, $ \begin{equation}
Z = -\frac{\pi \sigma^3}{6 V_k} \frac{\log(1-\epsilon)}{\Delta \varphi_k}
\end{equation}
where $ V_k $ is the volume at step $ k $ and $\Delta \varphi_k  =  (\varphi_{k+1} - \varphi_{k-1})/2 $ is the symmetric difference. 

\subsubsection{Dynamic Pressure Estimator}
Event chain Monte Carlo offers a direct method of calculating the equation of state, as described in Ref.\ \cite{Michel2014}.  Consider an event chain in the $+x$ direction.  At each collision between particles $ j $ and $ k $, the distance between the centers of the particles  projected on the direction of motion is $ x_k - x_j$.  Define the ``lifted" distance of an event chain,   $ x_\text{final} - x_\text{initial}$ as 
\begin{equation}
x_\text{final} - x_\text{initial} = \ell_c + \sum_{(k,j)} (x_k - x_j)
\end {equation}
where the sum is over all of the collision events in the chain.
The equation of state is then given by an average of the lifted distance over all event chains \cite{Michel2014},
\begin{equation} \label{eq:ecmc-pressure}
Z =  \bigg \langle \frac{x_\text{final} - x_\text{initial}}{\ell_c} \bigg \rangle_\text{chains}.
\end{equation} The results we present below for the equation of state are from this dynamic, ECMC estimator.

We  note that Eq.\ \eqref{eq:ecmc-pressure} is not a correct  measure of $Z$ when $ \ell_c > L$.  When  using the dynamic event chain length described in the Appendix, Eq.\ \eqref{eq:ecmc-pressure} cannot be used for $ \varphi \lesssim 0.15 $

\subsection{$\re$, $ \re^* $ and $\rt$}
In order to assess equilibration we computed the population size measures $ \re $, $ \re^* $, and $ \rt $ as described in Secs. \ref{sec:error} and \ref{sec:errort}. The mean square family size $ \rt $ is straightforward to measure in each run with Eq.\ \eqref{eq:rt} by keeping track of the family to which each replica belongs. These estimates can also be combined through a weighted average across independent runs.

To estimate the equilibrium population size $ \re $ through Eq.\ \eqref{eq:re}, however, we need the variance of entropy over many independent runs. However, because we use a variable annealing schedule, we do not have direct access to the variance of the entropy as a function of packing fraction. Instead we have variable packing fractions at each value of the entropy and $\text{var}(\varphi)$ as a function of $S_c$.
By considering the $S_c$ vs.\ $ \varphi $ curves from each run as independent random functions we can make the first-order estimate \begin{equation}
 \text{var}(\tilde{S}_c) = \bigg (\frac{\partial S_c}{\partial \varphi} \bigg)^2 \text{var}(\varphi),
\end{equation} with the same symmetric derivative estimator for $ \partial S_c / \partial \varphi $ as in Eq. \eqref{eq:pa-pressure}.

The situation is simpler for the weighted average equilibrium population size $ \re^*, $ because the weighted average entropy $ \bar{S}_c $ is calculated at fixed packing fractions, so $ \text{var}(\bar{S}_c) $ can be calculated directly by bootstrapping and Eq.\ \eqref{eq:re-star} gives $ \re^*. $

\section{Results}
\label{sec:results}

In this section we present the results of our 
simulations.     Our results address two objectives.  The first objective is to study the properties of the population annealing algorithm as applied, for the first time, to fluid systems and determine whether the algorithm can produce equilibrated results in the glassy regime of bidisperse hard spheres.  The second goal is to obtain precise equilibrated results for the equation of state at high densities and to study  nonequilibrium behavior at even higher densities near jamming for the 50/50 binary mixture of hard spheres with diameter ratio 1.4:1.   We study systems with $ N=60 $ and $ N=100$ particles.   For each of these cases we performed $M \approx 1000$ independent runs, each with population size $ R=10^6$. Table \ref{tab:trial-params} shows the simulation parameters.  Note that the population size (10$^6$) and the number ECMC sweeps per annealing step (one) are the same for both $N=60$ and 100 particles but the number of annealing steps required to keep the culling fraction fixed increases linearly in the number of particles.  This linear scaling is the result of the fact that the probability of an overlap for a given fractional compression increasing linearly in the increase in packing fraction.  Together with the linear scaling of carrying out a single sweep of  ECMC, we see that the naive complexity of the algorithm is $O(N^2)$.  Of course, this scaling does not take into account how computational resources must scale with $N$ to achieve equilibration.  The question of equilibration is discussed in the following subsection.   

\begin{table}

\caption{\label{tab:trial-params} Simulation parameters for the results reported in Sec.\ \ref{sec:results}.  $\varphi_{\rm max}$ is the maximum packing fraction.  Average wall clock time for a run corresponds to an  OpenMP implementation with 32 threads.
}

\begin{tabular}{| c | c | c | c |}
\hline
System Size $ N $ & & 60 & 100  \\ \hline
Population Size $ R $ & & $ 10^6 $ & $ 10^6 $ \\ \hline
Max packing fraction $\varphi_\text{max} $ & & 0.63 & 0.63 \\ \hline
Culling fraction $\epsilon$ && 0.05 & 0.05 \\ \hline
Avg \# Annealing Steps & & 11120 & 18800 \\ \hline
Average Time (hrs) & & 22 & 80 \\  \hline
Memory Usage (Mb) & & 1536 & 2452 \\ \hline
Number of Runs $ M $ & & 1069 & 986  \\ \hline
\end{tabular}

\end{table}

\subsection{Equilibration of Population Annealing}
\label{sec:paperf}

\begin{figure}
\centering
\includegraphics[width=0.7\linewidth]{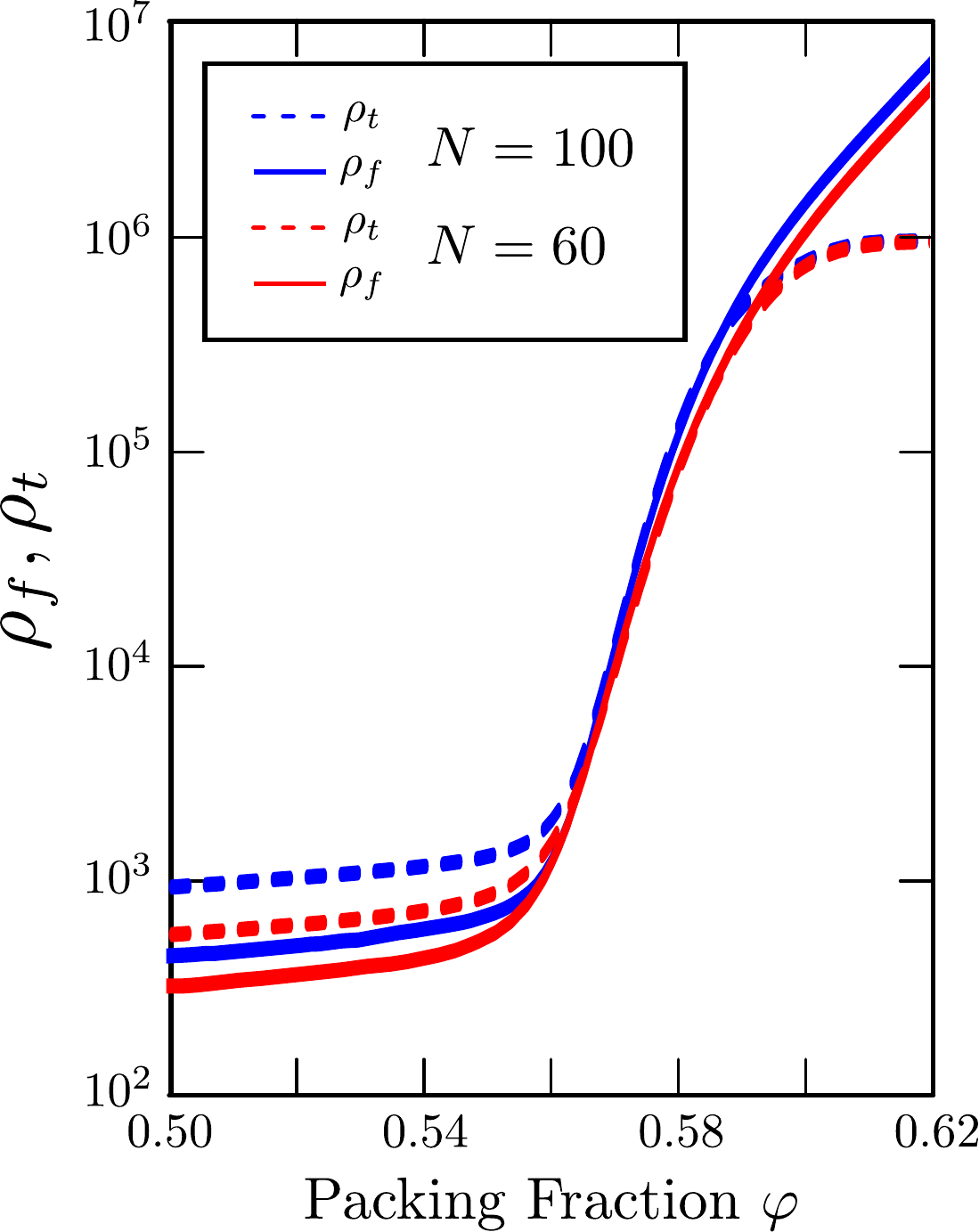}
\caption{(color online) Characteristic population sizes,  $\re$ (solid lines)  and $\rt$ (dotted lines) vs.\  packing fraction $\varphi$ for sizes $N=60$ (red) and $N=100$ (blue).  See Secs.\ \ref{sec:error} and \ref{sec:errort} for definitions.}
\label{fig:rho}
\end{figure}

Figure \ref{fig:rho} shows the characteristic population sizes $ \re $ and $ \rt$ vs.\ packing fraction.   These quantities are discussed in Secs.\ \ref{sec:error} and \ref{sec:errort}, respectively, and they characterize the errors in population annealing.  Specifically, when $\re$ is small compared to the population size, systematic errors are small and each run is well-equilibrated.  Similarly, when $\rt$ is small compared with the population size, statistical error are small. Figure \ref{fig:rho} demonstrates that both measures are relatively flat and much smaller than the population size ($R=10^6$) until $ \varphi = 0.55$, but then both $\re$ and $\rt$ grow quickly, increasing by about a factor of 100 between $ \varphi = 0.55 $ and 0.58.  If we impose a relatively conservative equilibration criterion that $R > 100 \re$ then individual runs are found to be well-equilibrated for $\varphi < 0.575$ for both system sizes.

Note that $\rt$ is approximately a factor of 2 larger than $\re$ at low density but in the region $0.56 \lesssim \varphi \lesssim 0.59$, the two quantities track each other closely (see Ref.\ \cite{toappear}).  Finally, for $\varphi \approx  0.59$ we find that $\re \approx \rt \approx R$.  Beyond this packing fraction, individual runs are clearly not in equilibrium.   Since $\rt$ is estimated from runs at population size $R$, it bounded by $R$ while $\re$ is  not bounded so the two quantities part ways above $\varphi \gtrsim 0.59$.

\begin{table}[h]
\caption{\label{tab:stat60}  Measured quantities at selected packing fractions, $\varphi$ for $N=60$ spheres.  $ \bar{S}_c$ is the weighted average entropy defined in Eq.\ \eqref{eq:weightedS} and $\bar{Z}$ is the weighted average equation of state with one standard deviation errors. $\mu$, $\sigma$,  and $\lambda$ are the parameters of the best fit exponentially modified Gaussian distribution (see text above Eq.\ \eqref{eq:exnormal}).  A dash in the $\lambda$ column indicates that a Gaussian distribution is the preferred fit.   $\rho_f$  and $\rho_f^*$ are the equilibration population sizes for single runs and  weighted averages defined in Eqs.\ \eqref{eq:re} and  \eqref{eq:re-star}, respectively. 
}
\begin{tabular}{|c|c|c|c|c|c|c|c|}
\hline
$ \varphi $ &$ \bar{S}_c$ & $\bar{Z}$  & $\mu$  & $\sigma$  & $\lambda$  & $\rho_f$  & $\rho_f^*$  \\ \hline
 0.54 & -343.71 & 16.2233(2) & -343.70 & 0.021 & - & 4.5$\times$ 10$^2$ & 4.5$\times$ 10$^2$ \\ \hline
 0.56 & -379.66 & 18.832(1) & -379.70 & 0.034 & - & 1.2$\times$ 10$^3$ & 1.2$\times$ 10$^3$ \\ \hline
 0.58 & -420.44 & 22.04(3) & -420.70 & 0.08 & 4.36 & 8.2$\times$ 10$^4$ & 3.2$\times$ 10$^5$ \\ \hline
 0.59 & -442.99 & 23.99(14) & -443.90 & 0.16 & 1.88 & 3.7$\times$ 10$^5$ & 8.2$\times$ 10$^6$ \\ \hline
 0.60 & -467.33 & 26.5  & -469.60 & 0.29 & 1.04 & 1.1$\times$ 10$^6$ & 9.4$\times$ 10$^7$ \\ \hline
\end{tabular}
\end{table}
\begin{table}[h]
\caption{\label{tab:stat100}  Measured quantities at selected packing fractions, $\varphi$ for $N=100$ spheres.  See Table \ref{tab:stat60} for details.}

\begin{tabular}{|c|c|c|c|c|c|c|c|}
\hline
$ \varphi $ &$ \bar{S}_c$ & $\bar{Z}$  & $\mu$  & $\sigma$  & $\lambda$  & $\rho_f$  & $\rho_f^*$  \\ \hline
 0.54 & -575.97 & 16.3455(1)  & -576.00 & 0.025 & - & 6.1$\times$ 10$^2$ & 6.1$\times$ 10$^2$ \\ \hline
 0.56 & -636.40 & 18.9956(7)  & -636.40 & 0.035 & - & 1.2$\times$ 10$^3$ & 1.2$\times$ 10$^3$ \\ \hline
 0.58 & -705.04 & 22.27(4)  & -705.40 & 0.10 & 3.68 & 1.2$\times$ 10$^5$ & 1.6$\times$ 10$^6$ \\ \hline
 0.59 & -743.19 & 24.51(6)  & -744.40 & 0.21 & 1.52 & 5.3$\times$ 10$^5$ & 1.6$\times$ 10$^7$ \\ \hline
 0.60 & -785.18 & 27.6  & -787.80 & 0.40 & 0.88 & 1.4$\times$ 10$^6$ & 5.8$\times$ 10$^7$ \\ \hline
\end{tabular}
\end{table}

Considering the ratios of $\re$ for the two system sizes it seems clear that in the glassy regime $\varphi \geq 0.58$ it is more difficult to equilibrate larger systems.  With only two system sizes, we cannot determine the scaling behavior of $\re$ with $N$.   

\begin{figure}
\centering
\includegraphics[width=0.8\linewidth]{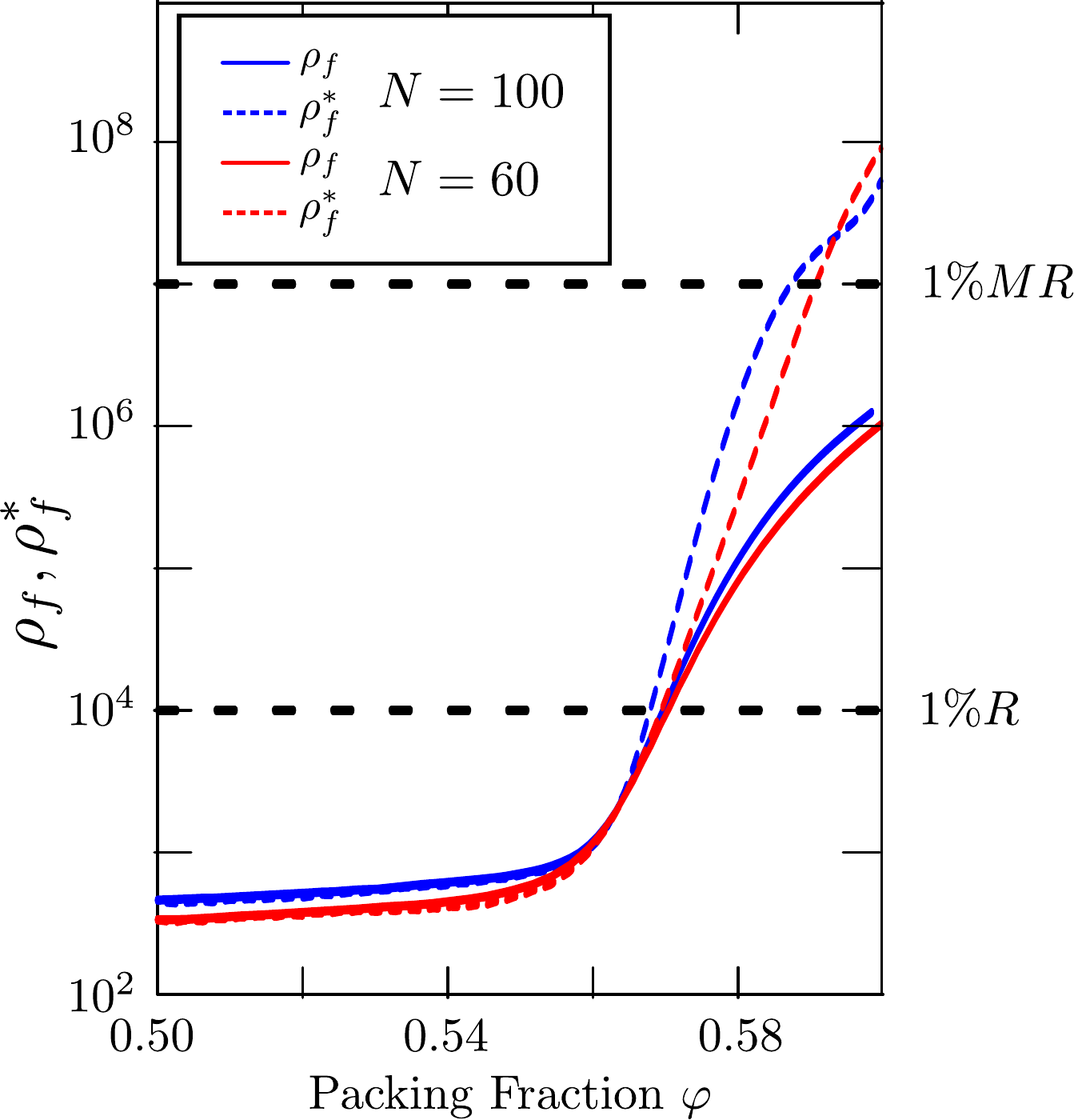}
\caption{(color online) Equilibration population sizes,  $\re$ (solid lines)  and $\re^*$ (dotted lines) vs.\  packing fraction $\varphi$ for sizes $N=60$ (red) and $N=100$ (blue).  See Sec.\ \ref{sec:error} for definitions. Dotted horizontal lines at $10^4$ and $10^7$ show where the cut-offs for equilibration occur for single runs and weighted averages, respectively,  based on the requirements $0.01 R >  \re$ for single runs and $0.01 MR >  \re^*$ for weighted averages.}
\label{fig:rho-star}
\end{figure}

Results from weighted averaging are equilibrated to higher densities than results from a single run.   Figure \ref{fig:rho-star} shows  $\re^*(R)$ the weighted average equilibration population size, together with $\re$, as a function packing fraction.  If we apply the  conservative equilibration criterion that $MR \geq 100 \re^*(R)$ then results from weighted averaging are well-equilibrated for $\varphi < 0.587$ for $N=100$ hard spheres and for $\varphi < 0.591$  for $N=60$ hard spheres.  Note that we believe that our results for the fluid equation of state remain reasonably accurate to somewhat higher packing fractions than these strict cut-offs.

\begin{figure}
\centering
\includegraphics[width=1.0\linewidth]{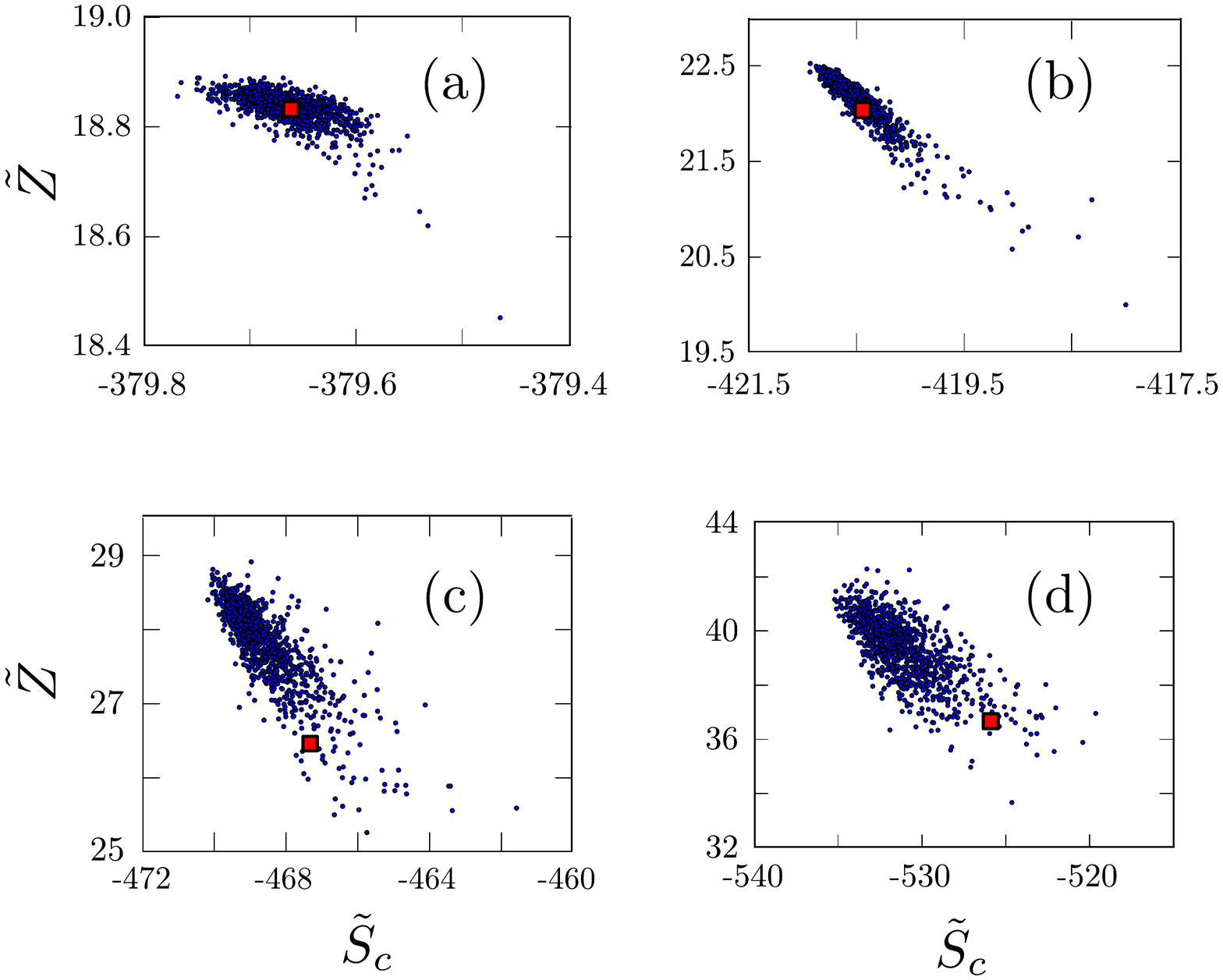}
\caption{(color online) Joint distribution of the entropy and equation of state estimators for $N=60$ and packing fractions $ \varphi= $ 0.56, 0.58, 0.60, and 0.62 for Figs. \ref{fig:pressure-entropy}a-d, respectively. Each point represents a single run of population annealing.  The $x$-coordinate of the point is the entropy estimator, $\tilde{S}_c$ and the $y$-coordinate is the equation of state estimator, $\tilde{Z}$. The weighted average values are shown as red squares. 
}
\label{fig:pressure-entropy}
\end{figure}

A deeper understanding of weighted averaging can be gained by looking at the joint distribution of the entropy and equation of state estimators, $\tilde{S}_c$ and $\tilde{Z}$, as a function of packing fraction.  Figure \ref{fig:pressure-entropy} displays these joint distributions as scatter plots for $N=60$ particles and several packing fractions.  Each point in these figures represents a single run of PA.  The horizontal coordinate of the point is the relative configurational entropy estimator $\tilde{S}_c$ and the vertical coordinate is the equation of state  estimator $\tilde{Z}$ for the run.  The $x$ and $y$ coordinates of the red squares are the weighted average values $\bar{S}_c$ and $\bar{Z}$, respectively.   Two important features of these plots are immediately evident: (1) there is an approximately inverse correlation between $\tilde{S}_c$ and $\tilde{Z}$ so that the weighted average value of the pressure is less than the ordinary average and (2) there are many outliers with large entropies and small pressures and the distributions are clearly not bi-variate Gaussians.    For $\varphi = 0.56$ and 0.58, these outliers have little effect on the weighted averages but for the $\varphi = 0.6$ and 0.62 they dominate the weighted averages.  For packing fractions less that 0.55, the distributions are well-described by bi-variate Gaussians.

\begin{figure}
\centering
\includegraphics[width=1\linewidth]{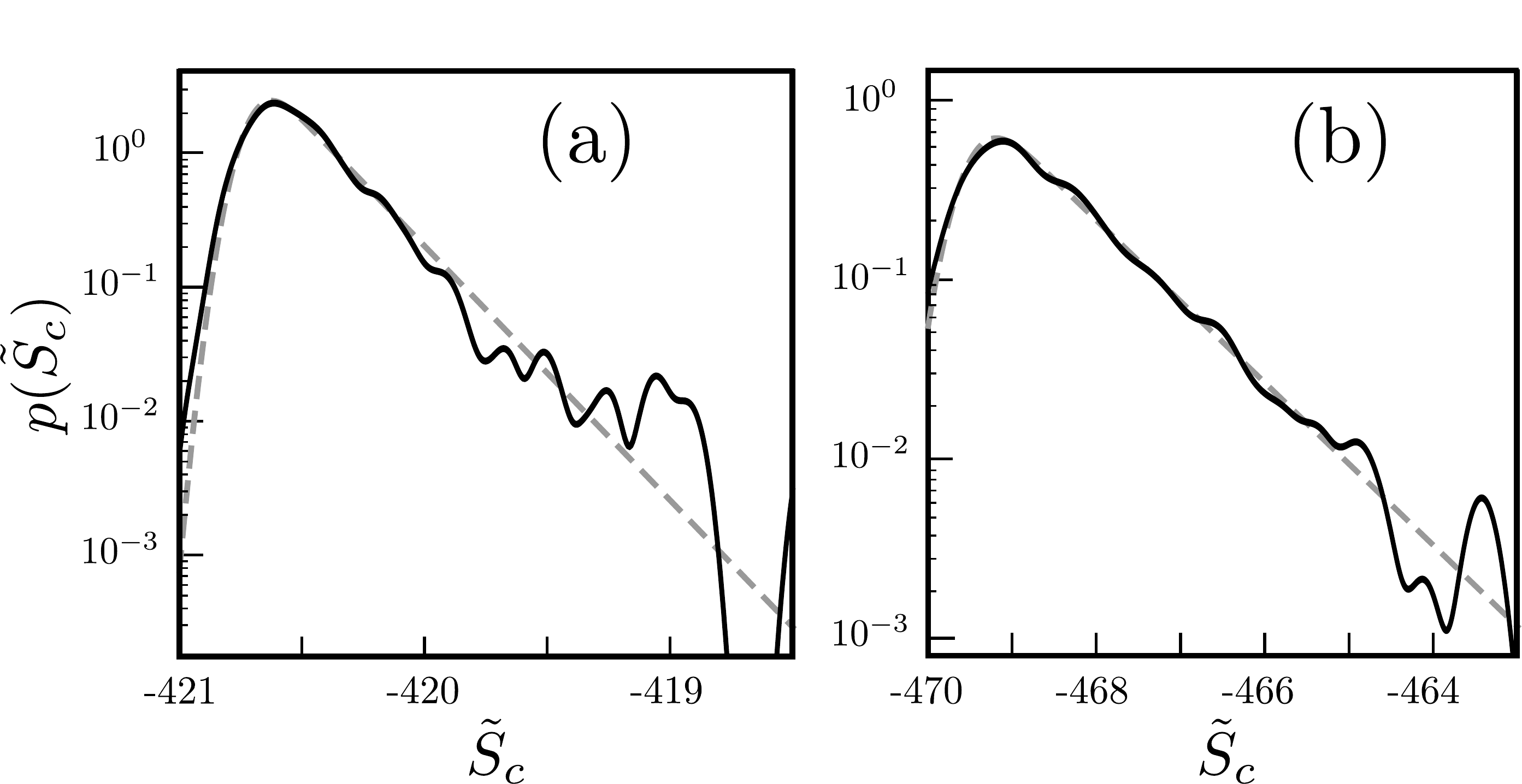}
\caption{ Distributions of the configurational entropy estimator $ \tilde{S}_c $ (solid curves) at packing fractions $\varphi=0.58$ (Fig. \ref{fig:exnormal-fit}a) and $\varphi=0.60$ (Fig. \ref{fig:exnormal-fit}b) together with a fit to an exponentially modified Gaussian distribution (dotted curves) for $N=60$ particles. See Table \ref{tab:stat60} for fitting parameters.}
\label{fig:exnormal-fit}
\end{figure}

The tail of the entropy distribution  determines the role of outliers in weighted averages.  Figure \ref{fig:exnormal-fit} shows logarithmic plots of the entropy estimator $\tilde{S}_c$ distributions (blue curves) for $\varphi =0.58$, and 0.60.  The distributions have been smoothed with Gaussian kernels using Mathematica's {\it KernelMixtureDistribution}.  It is clear that these distributions have exponential tails for large entropy rather than Gaussian tails.   A good fit to these distributions, shown as dotted curves in Fig.\ \ref{fig:exnormal-fit} , is obtained using an exponentially modified Gaussian, which is defined as follow:  Let $X$ be Gaussian distributed and $Y$ be exponentially distributed, then $Z=X+Y$ is described by an exponentially modified Gaussian distribution.  An exponentially modified Gaussian requires three parameters: $\mu$ and $\sigma$ are the mean and standard deviation of the Gaussian distribution, respectively,  and $\lambda$ is the characteristic decay rate of the exponential distribution.   The probability density function is given by
\begin{equation}
\label{ }
p(x; \mu,\sigma,\lambda)=\frac{\lambda}{2}   e^{\frac{1}{2} \lambda  \left(\lambda  \sigma ^2-2 x+2 \mu \right)} {\rm erfc}\left(\frac{\lambda  \sigma ^2-x+\mu
   }{\sqrt{2} \sigma }\right).
\end{equation}
The best fit values of the parameters for several values of $\varphi$ are shown in Tables \ref{tab:stat60} and \ref{tab:stat100}, for $N=60$ and 100 particles, respectively.  

If the entropy estimator is described by an exponentially modified Gaussian, the weighted average entropy $\bar{S}_c$, defined in Eq.\ \eqref{eq:weightedS},  approaches, as $M \rightarrow \infty$, the  integral of the distribution and the weighting factor,
\begin{equation}
\label{eq:exnormal}
\bar{S}_c = \log  \int dx \; p(x; \mu,\sigma,\lambda) e^x  .
\end{equation}
Carrying out the integral we find,
\begin{equation}
\label{ }
\bar{S}_c \rightarrow \mu + \frac{ \sigma^2}{2}  + \log \left(\frac{\lambda}{\lambda -1} \right) .
\end{equation}
When the exponential decay parameter $\lambda$ is significantly larger than one, the effect of the exponential tail is small and the weighted average is a good approximation for the entropy and other observables so long as the number of trials is sufficiently large to explore the tail, as is the case here with $M \approx 1000$.   However, if $\lambda \lesssim 1$ then the effect of the tail  on the weighted average diverges and features of the entropy distribution that have not yet been explored, for example,  a cut-off to the exponential tail, control the equilibrium averages.  In the high packing fraction region where,  $\lambda \lesssim 1$, the population annealing results do not yield useful information about the equilibrium properties of the system.  From Tables \ref{tab:stat60} and \ref{tab:stat100} we see that for $N=60$, $\lambda=1$ is reached at about $\varphi= 0.60$ while for $N=100$, $\lambda=1$ is reached between 0.59 and 0.60.  These estimates for where equilibration breaks down are consistent with the more conservative estimates based on $\re^*$.

\subsection{Equation of State}

\begin{figure}
\centering
\includegraphics[width=0.8\linewidth]{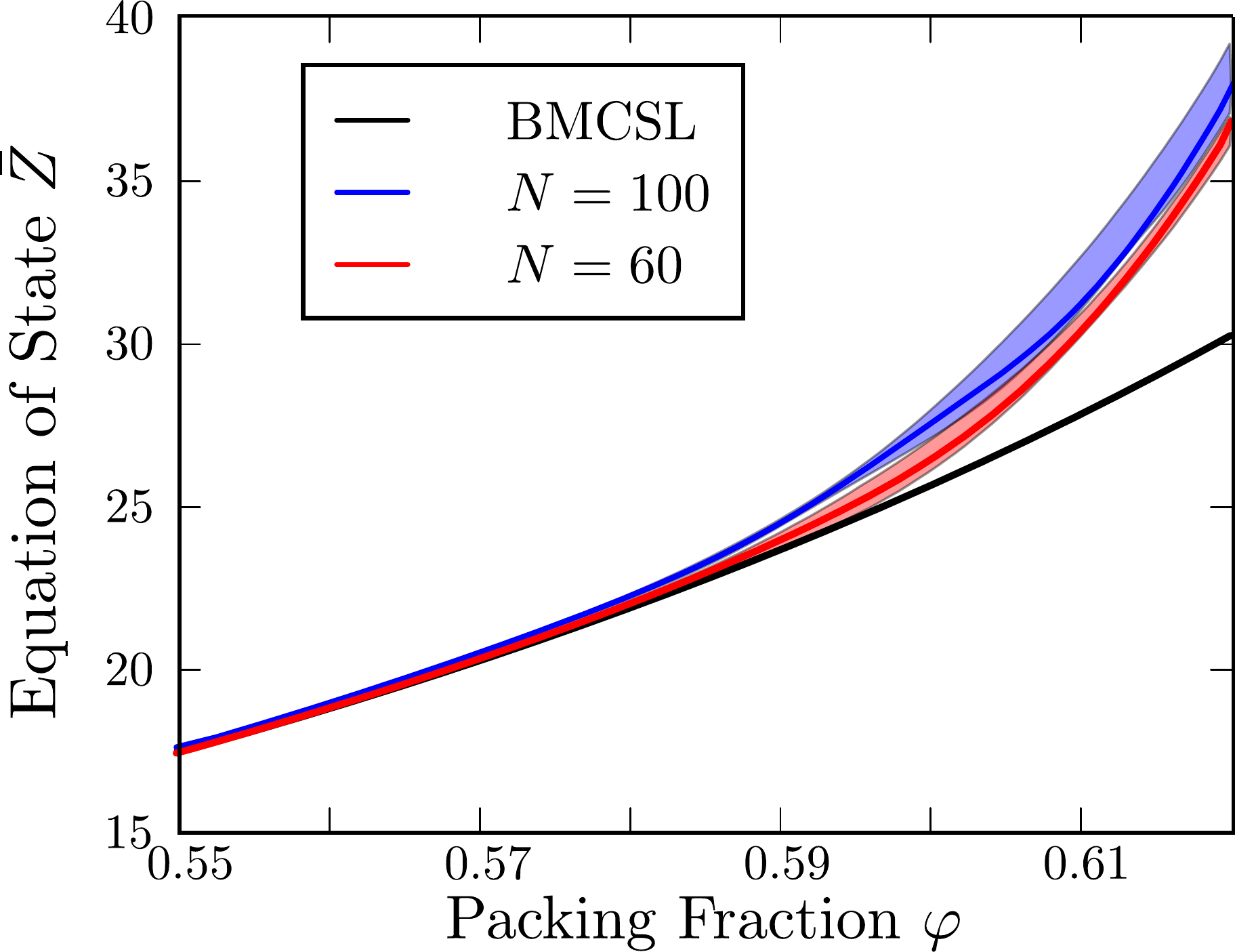}
\caption{(color online) The weighted average  equation of state, $\bar{Z}$ with bootstrapped 95\% confidence intervals for $N=60$ (red) and $N=100$ (blue) particles.  The solid black line is the BMCSL equation of state.}
\label{fig:eos}
\end{figure}

Figure \ref{fig:eos} shows the weighted average equation of state $\bar{Z}$ as a function of packing fraction $\varphi$ for systems with $ N=60 $ and $ N=100$ particles. The weighted average is performed using $ M \approx 1000 $ independent trials (see Table \ref{tab:trial-params}), each with population size $ R=10^6. $ We estimated the statistical errors by randomly resampling with replacement (bootstrapping) the collection of trials and recalculating the weighted average for the resampled trials.  We note that this method makes use of the joint distribution of entropy and pressure to estimate systematic errors in the equation of state.   The shaded regions around the weighted average curves represent  95\% confidence intervals. 
If the original weighted average is dominated by a small number of trials with significantly greater entropies, the bootstrapped collection may leave out these trials, resulting in a highly skewed confidence interval. This effect is clearly seen in Fig. \ref{fig:eos} at $ \varphi \approx 0.61$ where the simulations have clearly fallen out of equilibrium.

\begin{figure}
\centering
\includegraphics[width=0.8\linewidth]{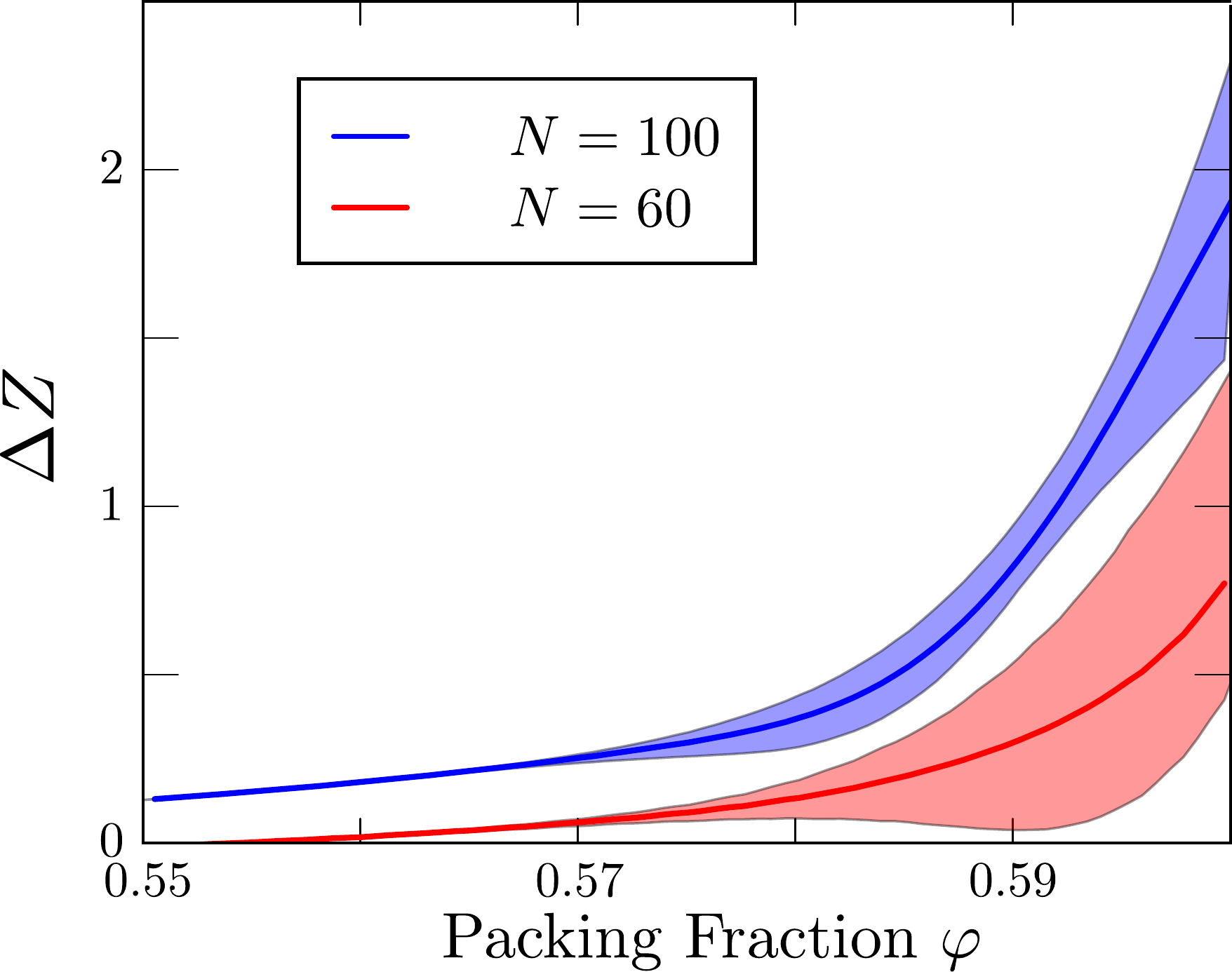}
\caption{(color online) Deviations $\Delta Z$ between the simulation results and the BMCSL equation of state as a function of packing fraction $\varphi$ with bootstrapped 95\% confidence intervals for $N=60$ (red) and $N=100$ (blue) particles.}
\label{fig:bmcsl-deviation}
\end{figure}

The solid line in Fig.\ \ref{fig:eos} is the BMCSL equation of state (see Eq.\ \eqref{eq:bmcsl}), which is reasonably accurate for low densities but significantly underestimates $Z$ for high densities. Of course, the BMCSL equation predicts finite pressure up to $ \varphi = 1$  so that deviations of this kind are inevitable but it is not clear where they become significant.  Figure \ref{fig:bmcsl-deviation} shows the difference $ \Delta Z = \bar{Z} - Z_{\mathrm{CS}} $ between the equation of state measured in the simulations and the BMCSL prediction. For $N=60$, significant deviations begin at $ \varphi = 0.56$ while for $N=100$ small deviations persist to much lower packing fractions.  Large differences from BMCSL begin at $\varphi=0.58$.     

\begin{figure}
\centering
\includegraphics[width=0.8\linewidth]{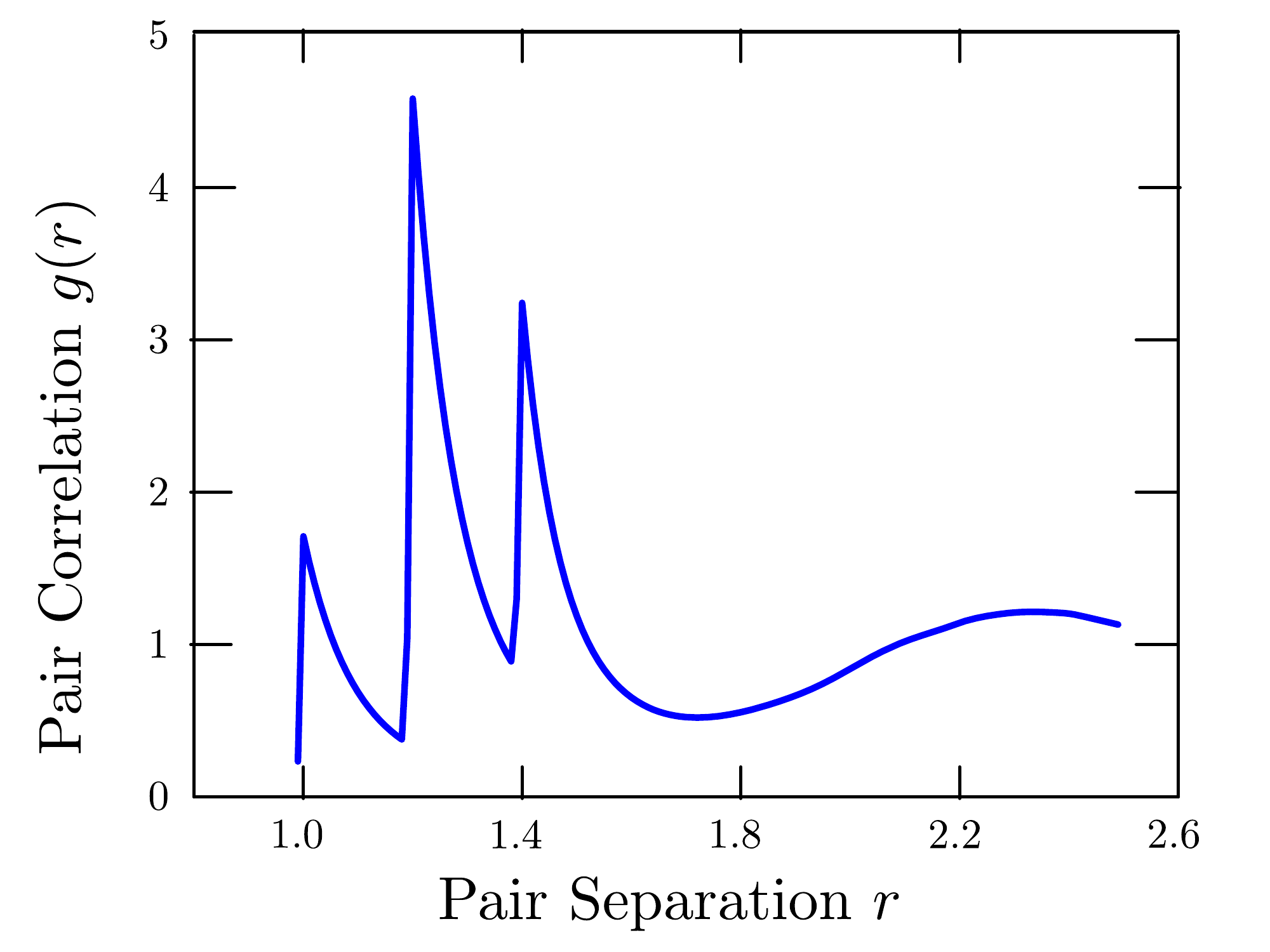}
\caption{Pair correlation function $g(r)$ for $N=100$ particles at packing fraction $ \varphi = 0.58 $ as a function of separation $r$. 
The three peaks correspond to the separations at contact for this binary mixture.
}
\label{fig:pair-correlation}
\end{figure}

Since we are interested in the (metastable) equilibrium behavior of the  fluid and not possible ordered states, we checked for ordering with the pair correlation function $g(r)$.  The pair correlation function is shown in Fig.\ \ref{fig:pair-correlation} for $N=100$ particles at packing fraction $ \varphi=0.58 $.  The pair correlation function for $N=60$ particles is indistinguishable from $g(r)$ for $N=100$ particles.
We see sharp peaks only at the three contact distances for this binary mixture.   If the system typically formed phase-separated crystals or other ordered states we would expect additional sharp structure in $g(r)$.   In addition, Fig.\ \ref{fig:pair-correlation} is obtained from a weighted average over runs but we observed no noticeable difference between the weighted and unweighted averages.  This result suggests that the entropy of these systems is not strongly correlated with the $g(r)$.  Although we cannot rule out the possibility that a small fraction of configurations in the ensemble are ordered, we conclude that the predominant configurations sampled by the algorithm are amorphous.

\begin{figure}
\centering
\includegraphics[width=0.8\linewidth]{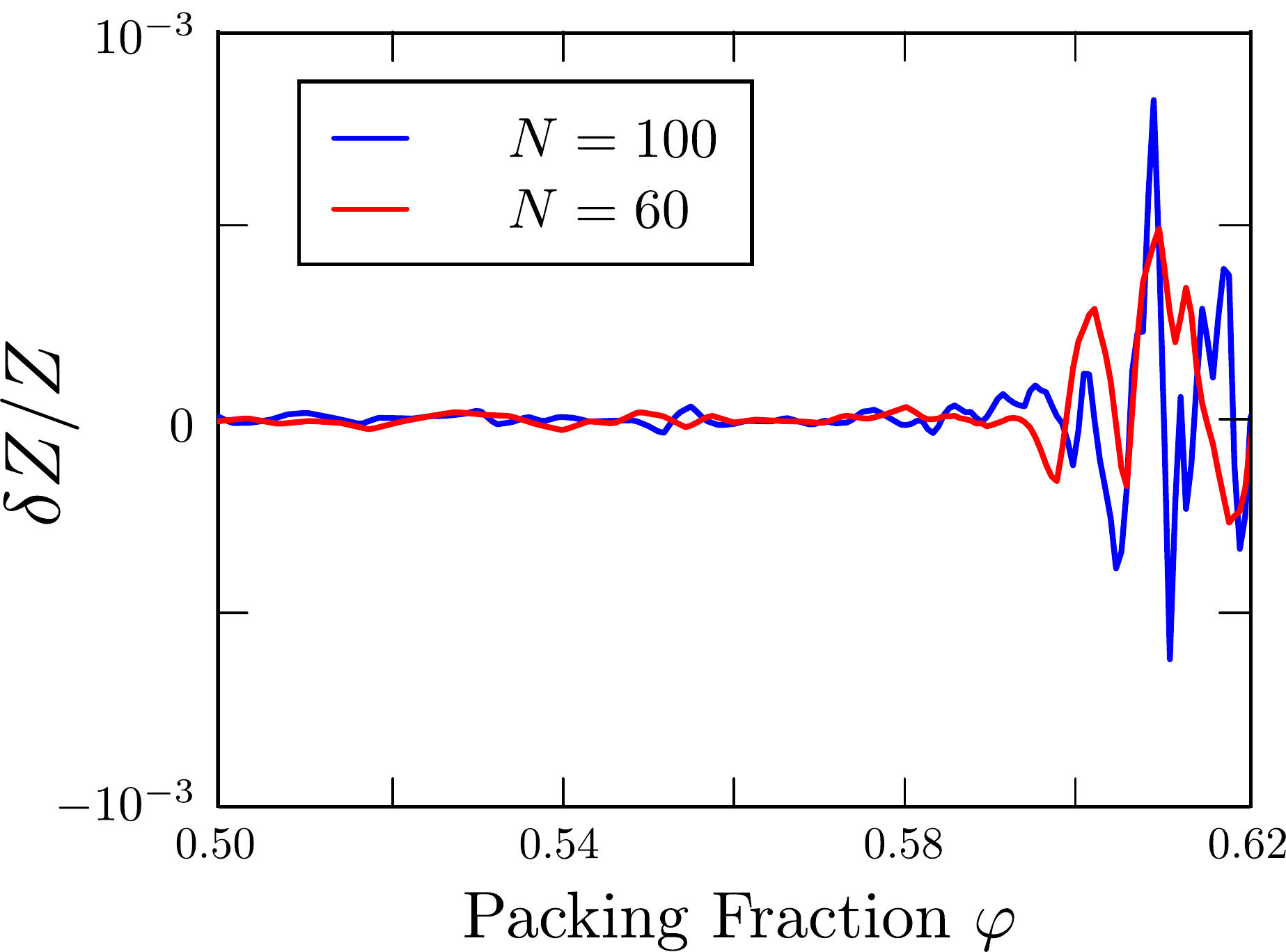}
\caption{(color online) Relative difference $\delta Z/Z$ between the dynamic  and thermodynamic  estimates of the equation of state, Eqs \eqref{eq:ecmc-pressure} and \eqref{eq:pa-pressure}, respectively, for $N=60$ (red) and $N=100$ (blue) particles.}
\label{fig:pa-vs-ecmc}
\end{figure}

The above results for the equation of state have been obtained using the dynamic measure of the pressure.  Figure  \ref{fig:pa-vs-ecmc}, shows the relative discrepancy, $\delta Z/Z$ between the thermodynamic and dynamic measures of pressure, given in Eqs.\ \eqref{eq:pa-pressure} and \eqref{eq:ecmc-pressure}, respectively. We find that  $ \delta Z / Z  \approx 10^{-4}$ until $ \varphi \approx 0.59$. The good agreement between these two independent measures is an important validation of the algorithm and gives us confidence that estimates of the equation of state based on Eq.\ \eqref{eq:pa-pressure} are quite accurate.

\subsection{High Density Behavior}

\begin{figure}
\centering
\includegraphics[width=0.8\linewidth]{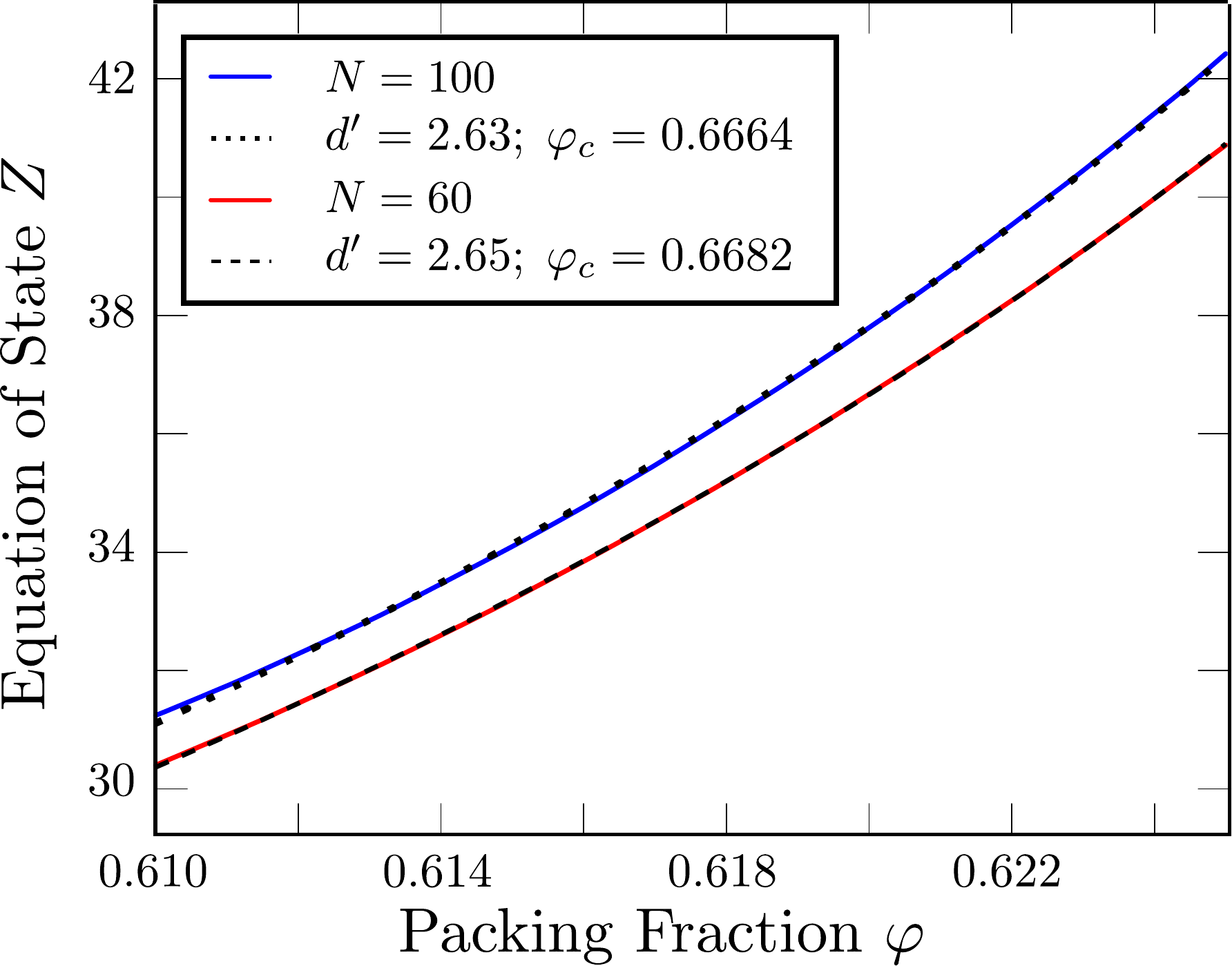}
\caption{High density non-equilibrium equation of state, $\bar{Z}$ (solid curves) along with fits (dotted curves) to the free volume form, Eq.\ \eqref{eq:freevolume} for $N=60$ (red) and $N=100$ (blue) particles.}
\label{fig:pole-divergence}
\end{figure}

Though the simulations have fallen out of equilibrium for $ \varphi \gtrsim 0.6$, population annealing continues to work as a nonequilibrium protocol to achieve high density packings and nearly jammed states.  We fit our equation of state data in the range   $0.61 <\varphi< 0.625$ to the ``free volume'' form,
\begin{equation}
\label{eq:freevolume}
Z= \frac{d' \varphi_c}{\varphi_c - \varphi}.
\end{equation}
Table \ref{tab:free-vol} gives the parameters of the fit and Fig. \ref{fig:pole-divergence} shows the simulation data along with fits.
It should be noted that in this range of packing fractions the weighted average is dominated by the single highest entropy run so that it is likely that the estimate for $\varphi_c$ would increase as the number of runs or population size increases.  It would be interesting to study how $\varphi_c$ changes as either population size or number of runs increases.
Our results agree reasonably well with Odriozola and Berthier~\cite{Odriozola2011} who find $ d' = 2.82$ and  $\varphi_c = 0.669$, from fits in the range $0.61 < \varphi \lesssim 0.65$ . 

\begin{table}
\caption{\label{tab:free-vol}  Parameters and bootstrapped 95\% confidence interval for fitting the equation of state to the free volume form of Eq. \eqref{eq:freevolume} for the range of $ 0.61< \varphi <0.625$.}

\begin{tabular}{|c|c|c|c|}
\hline
System Size $ N $ & & $ d' $ & $ \varphi_c $  \\ \hline
 60 & & $ 2.65 \substack{+0.24 \\ -0.11} $ & $0.668 \substack{+0.005 \\ -0.002}$  \\ \hline
 100 & & $ 2.63 \substack{+0.62 \\ -0.08} $ & $0.666 \substack{+0.011 \\ -0.002} $ \\ \hline
\end{tabular}
\end{table}

To explore the behavior of population annealing as a  protocol for achieving nearly jammed states, we  allowed ten runs to go to much higher packing fraction. These runs terminated after exceeding a maximum run time and reached packing fractions in the range $0.660<  \varphi_c < 0.663$. We fit each run independently to Eq. \eqref{eq:freevolume} and obtained $\varphi_c$ in the range $0.662 <  \varphi_c < 0.664$. These values of $\varphi_c$ are smaller than the estimates shown in Table  \ref{tab:free-vol} because they represent only 10 rather than 1000 independent experiments.

\section{Discussion}
\label{sec:discussion}

We have developed  population annealing Monte Carlo for fluid systems and applied it to a glass-forming binary mixture of hard spheres.  We find that population annealing is a promising method for computational studies of fluids in the high density regime.  It is a highly parallelized algorithm that is well-suited to simulations on large computer clusters either by using weighted averaging and a large number of independent runs, as we have done here, or by carrying out a smaller number of very large population runs with a massively parallelized implementation of the algorithm.  The advantage of the former approach is that we can learn much about the equilibration of weighted averages from the statistics of multiple runs but this advantage is balanced by the lesser efficiency of weighted averaging as opposed to a single large population run.

We measured the equation of state in the glassy region of a 50/50 mixture of hard spheres with diameter ratio 1.4:1 and have obtained  precise results that are in reasonable agreement with previous simulations using parallel tempering~\cite{Odriozola2011}.  We find good agreement with the BMCSL equation of state up to a packing fraction of 0.58 but strong deviations above that packing fraction.  Although this is also the region where equilibration becomes much for difficult, we believe the simulations are reasonably well equilibrated up to a packing fraction of 0.60.    We also studied the equation of state at higher densities where population annealing serves as a non-equilibrium protocol for generating nearly jammed states. 

We must emphasize that our results at high densities likely describe the metastable equilibrium fluid branch of the equation of state and not the true equilibrium, which presumably consists either of a phase separated crystalline state or, for small $N$, perhaps a non-random best packing of $N$ spheres.  The definition of metastable equilibrium depends on the protocol used to sample the distribution so, in the region where the true equilibrium ensemble contains a significant contribution from non-random states, our results may differ from those obtained from other algorithms.  It would be interesting to investigate the true equilibrium states of the binary mixture studied here for finite $N$.

This work is the first application of population annealing to classical fluids and there are undoubtably many avenues for improvement. We have implemented the algorithm in the NVT ensemble but it is important to study population annealing in the NPT ensemble.  It would also be interesting to explore other annealing schedules.  For example, a variable number of Monte Carlo sweeps with more sweeps in the region of the glass transition may improve efficiency.

\appendix
\section{Dynamic chain length}
\label{app:chain-length}

Based on the average number of particles displaced in each event chain, we define an ECMC sweep to be a number of event chain moves that will move approximately $ N $ particles on average. We arbitrarily choose to divide this computational work so that an average of $ \sqrt{N} $ particles are displaced in each event chain. Then $ \sqrt{N} $ event chains constitutes an ECMC sweep of the system.

The event chain length, $ \ell_c$ must be chosen dynamically,  which can be done using Eq.\ \eqref{eq:ecmc-pressure}, re-written in the form, \begin{equation}
Z = 1 + \frac{\big \langle \sum_{\text{chains}} (x_k - x_j) \big \rangle_\text{chains}}{\ell_c}.
\end{equation}
The ``lifting distance" $ x_j - x_i $ between two particles of diameter $ \sigma $ with bond orientation $ (\theta, \varphi) $ at contact is 
\begin{equation}
x_j - x_i = \sigma \sin \theta \cos \varphi.
\end{equation} In a fluid, this bond orientation should be random, so the average projected distance is \begin{equation}
\big \langle x_j - x_i \big \rangle = \frac{1}{2 \pi} \int_{0}^{\pi} \int_{-\pi/2}^{\pi/2} d\Omega \, \sigma \sin \theta \cos \varphi = \frac{\sigma}{2}.
\end{equation} 
This will be somewhat less accurate in an fcc solid, but it turns out to be close enough to give a reasonable estimate. 
If an event chain consists of $ n_c $ collisions, 
\begin{equation}
\big \langle \sum_{\text{chains}} (x_j - x_i) \big \rangle_\text{chains} \approx n_c \sigma / 2 ,
\end{equation} 
and we find, 
\begin{equation}
Z \approx \frac{\ell_c + n_c \sigma / 2}{\ell_c}.
\end{equation} 
We would like $ \bar{n}_c \approx \sqrt{N}$.  For density $ \varphi_i $ then we use the equation of state $ Z (\varphi_{i-1}) $ at the previous density to estimate $ \ell_c (\varphi_i) $ as, 
\begin{equation}
\ell_c (\varphi_i) \approx \frac{ \sigma \sqrt{N} \,  / 2}{Z(\varphi_{i-1}) - 1}.
\end{equation} 
Empirically, we found a somewhat better approximation to $n_c = \sqrt{N}$ using 
 \begin{equation}
\ell_c (\varphi_i) = \frac{\sigma \sqrt{N/ 2}}{Z(\varphi_{i-1}) - 1} .
\end{equation}

\acknowledgments
This work was supported by the National Science Foundation (Grant No.~DMR-1507506).  We thank the Massachusetts Green High Performance Computing Center (MGHPCC) and the University of Massachusetts Amherst for providing computing resources.

\end{document}